\documentclass[prl,twocolumn,showpacs,preprintnumbers,aps,nofootinbib]{revtex4-1}
\usepackage[dvipdfmx]{graphicx}
\usepackage{dcolumn}
\usepackage{bm}
\usepackage{amsmath,amssymb,amsfonts}
\usepackage{latexsym}
\usepackage{bbm}
\usepackage[dvipdfmx]{color} 
\begin{document}

\preprint{ACFI-T16-34}

\title{Does a band structure affect sphaleron processes?}

\author{Koichi Funakubo$^1$}
\email{funakubo@cc.saga-u.ac.jp}
\author{Kaori Fuyuto$^{2}$}
\email{kfuyuto@umass.edu}
\author{Eibun Senaha$^3$}
\email{eibunsenaha@ntu.edu.tw}
\affiliation{$^1$Department of Physics, Saga University, Saga 840-8502 Japan and}
\affiliation{$^2$Amherst Center for Fundamental Interactions, Department of Physics, University of Massachusetts Amherst, MA 01003, USA}
\affiliation{$^3$Department of Physics, National Taiwan University, Taipei 10617, Taiwan}
\bigskip

\date{\today}

\begin{abstract}
Inspired by a recent work of Tye and Wong, we examine an effect of a band structure 
on baryon number preservation criteria requisite for successful electroweak baryogenesis.
Action of a reduced model is fully constructed including a time component of the gauge field that is missing in the original work. 
The band structure is estimated more precisely in wider energy range based on WKB framework with three connection formulas to find
that the band structure has little effect on the criteria at around 100 GeV temperature.
We also address an issue of suppression factors peculiar to the $(B+L)$-changing process in high-energy collisions at zero temperature.

\end{abstract}


\maketitle

\paragraph{Introduction.---}
It is known that conservation of baryon $(B)$ and lepton number $(L)$ is nonperturbatively violated 
by the chiral anomaly in electroweak theories~\cite{'tHooft:1976up}. 
Experimental verification of such a violation is important not only for understanding
the nonperturbative nature of the quantum field theory but also for phenomenology in particle physics
and cosmology.
In particular, existence of such an anomalous process has expanded the possibilities of baryogenesis
in the early Universe: electroweak baryogenesis (EWBG)~\cite{ewbg}, leptogenesis~\cite{Fukugita:1986hr}, 
and baryogenesis via neutrino oscillations~\cite{Asaka:2005pn}, etc. 

It is shown by 't~Hooft that the $(B+L)$-changing vacuum transition 
is suppressed by $e^{-S_{\rm instanton}}\simeq 10^{-162}$, where $S_{\rm instanton}$ is the instanton action. Later, the analysis was generalized to the process in high-energy collisions
and found that the above exponential suppression can get alleviated with increasing energy
~\cite{Ringwald:1989ee, Espinosa:1989qn}. 
However, the perturbative calculation using an instanton-like configuration 
is no longer valid as the energy approaches to a sphaleron energy ($E_{\rm sph}$)
which corresponds to the height of the barrier separating topologically 
different vacua. A nonperturbative analysis in a simplified model indicates 
that the exponential suppression still persists even at very high energy~\cite{Son:1993bz}.

Recently, Tye and Wong revisited the possibility of the $(B+L)$-changing process at high energies
in the framework of a reduced model~\cite{Tye:2015tva}. 
In their analysis, the reduced model is quantized so that the energy spectrum has a band structure
owing to the periodic potential, and the wave functions are given by Bloch waves. 
It is demonstrated that the band width gets wider with increasing energy, 
and the energy spectrum becomes almost continuous above $E_{\rm sph}$, and therefore the transition amplitude is free from the exponential suppression
when the energy is larger than $E_{\rm sph}$. 
Such a description is missing in the past work.
After their claim, discussions on the detectability of the $(B+L)$-changing process at colliders or
in high-energy cosmic rays have been revived~\cite{BandSph_detection}. 

While the $(B+L)$-changing process is unsuppressed at higher temperature than the electroweak phase transition,
its probability below that temperature should be so small to satisfy the baryon-number preservation criteria (BNPC) for a successful EWBG scenario.
It is a natural question to what extent the above band structure description can change the BNPC. 
Such an analysis is enormously important for a test of EWBG at the Large Hadron Collider (LHC)
since the BNPC determines the minimal value of strength of the first-order electroweak phase transition,
so the collider signatures in the Higgs sector.

In this Letter, we investigate the effect of the band structure on the BNPC.
As is done in Refs.~\cite{Aoyama:1987nd,Funakubo:1990rz,Funakubo:1991hm,Funakubo:1992nq,Tye:2015tva}, the reduced model is constructed by regarding the noncontractible loop parameter~\cite{sph}
as a dynamical variable.
Unlike the work of~\cite{Aoyama:1987nd,Tye:2015tva}, we adopt a fully gauge-invariant 
approach in which the time component of the gauge field $A_0$ is also taken into account
for the construction.
Furthermore, we evaluate the band structure in a more precise manner utilizing the WKB method 
with three connection formulas depending on energy. With those improvements, we compare our
results with those in Ref.~\cite{Tye:2015tva} and quantify the impact of the band structure on EWBG.

We also point out a problem of an exponential suppression factor besides the tunneling factor
in the $(B+L)$-changing process in high-energy collisions at zero temperature,
which is not properly argued in Ref.~\cite{Tye:2015tva}.

\paragraph{The reduced model.---}
\label{model}
We consider the standard model (SM) for an illustrative purpose. 
The generalization to models beyond the SM is straightforward. 
Since the effect of a $U(1)_Y$ contribution on the sphaleron energy is a few \%, we neglect it. 
The starting point is the following $SU(2)_L$ gauge-Higgs system:
\begin{align}
{\cal L}=-\frac{1}{4}F^a_{\mu\nu}F^{a\mu\nu}+(D_{\mu}\Phi)^{\dagger}D^{\mu}\Phi
-\lambda\left(\Phi^{\dagger}\Phi-\frac{v^2}{2}\right)^2,
\end{align}
where $D_{\mu}=\partial_{\mu}+ig_2A_{\mu}$ with $g_2$ being the gauge coupling.
$F^{a}_{\mu\nu}$ is the field strength tensor and $\Phi$ the $SU(2)_L$ doublet Higgs field. 
The vacuum expectation value  of the Higgs field is denoted as $v(\simeq246~{\rm GeV})$ .
With the Manton's ansatz~\cite{sph}, the gauge and Higgs fields are cast into the form
\begin{align}
 A_i(\mu,r,\theta,\phi)=&\frac{i}{g_2}f(r)\partial_iU(\mu,\theta,\phi)U^{-1}(\mu,\theta,\phi), \label{gauge_ansatz}\\
{\Phi}(\mu,r,\theta,\phi)=&\frac{v}{\sqrt{2}}\bigg[(1-h(r))
\begin{pmatrix}
0\\
e^{-i\mu}\cos\mu
\end{pmatrix}\nonumber\\
&\hspace{1cm}
+h(r)U(\mu,\theta, \phi)
\begin{pmatrix}
0\\
1
\end{pmatrix}
 \bigg],\label{higgs_ansatz} 
\end{align}
with
\begin{align}
&U(\mu,\theta,\phi)\nonumber\\
&=
\begin{pmatrix}
e^{i\mu}(\cos\mu-i\sin\mu\cos\theta) & e^{i\phi}\sin\mu\sin\theta\\
-e^{-i\phi}\sin\mu\sin\theta & e^{-i\mu}(\cos\mu+i\sin\mu\cos\theta),
\end{pmatrix}
\end{align}
with a noncontractible loop parameter $\mu$ which runs from 0 to $\pi$. 
Note that Eqs.~(\ref{gauge_ansatz}) and (\ref{higgs_ansatz}) are reduced to the vacuum configurations
for $\mu=0$ and $\pi$ while the sphaleron for $\mu=\pi/2$.

Our aim is to study the transition from one vacuum to another along the least energy path 
such that the sphaleron configuration is realized at the maximal point.
For this purpose, $\mu$ is promoted to the dynamical value $\mu(t)$
as proposed in Ref.~\cite{Aoyama:1987nd}.
In contrast to the previous studies~\cite{Aoyama:1987nd,Tye:2015tva}, 
we construct the reduced model in a fully gauge-invariant way, where $A_0=ifU^{-1}\partial_0U/g_2$.
Note that the Manton's ansatz with the $A_0=0$ gauge causes 
an unwanted divergence arising from the $D\Phi$ term in asymptotic region $r\to\infty$
due to lack of the full gauge invariance, 
and therefore some prescriptions are needed~\cite{Aoyama:1987nd,Tye:2015tva}. 

After deriving the profile functions $f(r)$ and $h(r)$ by solving the equations of motion for the sphaleron, one obtains the classical action as
\begin{align}
S[\mu]=g_2v\int dt~\bigg[\frac{M(\mu)}{2}\left(\frac{d}{dt}\frac{\mu(t)}{g_2v}\right)^2-V(\mu)\bigg],
\label{S_mu}
\end{align}
where
\begin{align}
M(\mu)&=\frac{4\pi}{g^2_2}\left(\alpha_0+\alpha_1\cos^2\mu+\alpha_2\cos^4\mu\right),\\
V(\mu)&=\frac{4\pi}{g^2_2}\sin^2\mu \left(\beta_1+\beta_2\sin^2\mu\right).
\end{align}
The coefficients of $\alpha$'s and $\beta$'s are found to be
\begin{align}
\alpha_0&=19.42,\quad \alpha_1=-1.937, \quad \alpha_2=-2.656,\nonumber\\
\beta_1&=1.313,\quad \beta_2=0.603,
\end{align}
The sphaleron mass and potential in units of mass are, respectively, given by
\begin{align}
M_{\rm sph} &= g_2vM\left(\frac{\pi}{2}\right) \simeq 92.0~{\rm TeV}, \\
E_{\rm sph}&=g_2vV\left(\frac{\pi}{2}\right)\simeq 9.08~{\rm TeV}.
\end{align}
Note that $M_{\rm sph}$ depends on the normalization of the kinetic term in Eq.~(\ref{S_mu})
which is different from that in Ref.~\cite{Tye:2015tva} by a factor of 4.
With the same normalization, $M_{\rm sph}=23.0$ TeV in our case
and $M_{\rm sph}=17.1$ TeV in Ref.~\cite{Tye:2015tva},
which implies that the $A_0$ contribution to $M_{\rm sph}$ is not negligible.
As will be discussed below, number of the bands can change according 
to the size of $M_{\rm sph}$.

When a classical Hamiltonian is quantized, 
an ambiguity arises from an operator ordering. 
In our analysis, we adopt
\begin{align}
\mathcal{H}(\mu,p)&=g_2v\left[\hat{p}\frac{1}{2M(\hat{\mu})}\hat{p}+V(\hat{\mu}) \right],
\end{align}
where $\hat{p}$ is the momentum conjugate operator that satisfies $[\hat{\mu},~\hat{p}]=i$.
With this Hamiltonian, we solve the Schr\"odinger equation ${\cal H}\psi(\mu)={\cal E}\psi(\mu)$,
where ${\cal E}=E/g_2v$. 
Since the potential is a periodic function of $\mu$, 
the eigenfunction $\psi(\mu)$ is given by Bloch wave and energy spectrum 
has the band structure.

The band edges are determined, in the WKB approximation, as solutions to
\begin{align}
\cos(\Phi({\cal E}))=\pm \sqrt{T({\cal E})}, \label{band_edge}
\end{align}
where $T({\cal E})$ is a transmission coefficient at ${\cal E}$ and $\Phi({\cal E})$ is defined by
\begin{equation}
  \Phi({\cal E}) = \int^{a({\cal E})}_{b({\cal E})}d\mu\,\sqrt{2M(\mu)({\cal E}-V(\mu))},
\end{equation}
where $a({\cal E})$ and $b({\cal E})$ are the turning points determined by ${\cal E}=V(\mu)$ for ${\cal E}<V(\pi/2)$, while
$a({\cal E})=\pi/2$ and $b({\cal E})=-\pi/2$ for ${\cal E}\ge V(\pi/2)$~\cite{Balazs}.
We denote the lower and upper edges of the $n$-th band as ${\cal E}_{-,n}$ and ${\cal E}_{+,n}$, respectively, so its band width is given by
$\Delta {\cal E}_n={\cal E}_{+,n}-{\cal E}_{-,n}$.
Depending on the energy, three kinds of connection formulas are properly used:
linear potential, parabolic potential and over-barrier approximations 
(for a review, see, {\it e.g.}, Ref.~\cite{Landau:1977}).
Detailed calculations will be given in Ref.~\cite{FFS}.
To verify the validity of this method, 
we also applied it to an eigenvalue problem with a sine-type potential, 
where the band edges are known to be the characteristic values of the Mathieu function, 
and then confirmed that the above method sufficiently works well.

Table~\ref{band_structure} shows the band structure in the reduced model. 
The number of the bands below $E_{\rm sph}$ is 158 in our case while 148 
in Ref.~\cite{Tye:2015tva}, which is attributed to the different values of $M_{\rm sph}$.
While the band center energies are more or less the same as those in Ref.~\cite{Tye:2015tva}, significant differences exist in the band widths at $E\simeq \mathcal{O}(100)$ GeV. 
However, such a difference does not give any impact on the results we are concern with.

One can see that the band widths become wider as the energy grows, while the band structure persists above $E_{\rm sph}$ even though the band
gaps gets smaller. 



\begin{table}[t]
  \begin{tabular}{|c |c c |}
  \hline
  $n$ & Lower edge [GeV] & Band width [GeV]\\
  \hline
  \hline
1&  $33.827$ & $1.304\times10^{-199}$\\
2 &  $101.466$ & $1.879\times10^{-196}$\\
3 &  $169.080$ &  $1.309\times10^{-193}$\\
4&  $236.667$ & $6.044\times10^{-191}$\\
5 &  $304.224$ & $2.09\times10^{-188}$\\
6 &  $371.751$ &  $5.776\times10^{-186}$\\
$\vdots$  & $\vdots$ &  $\vdots$\\
158&  $9066.503$ &  $10.355$\\
159  & $9087.266$ &  $19.205$\\
160 & $9112.052$ &  $27.172$\\
161  & $9141.122$ &  $31.580$\\
162 & $9173.272$ & $34.027$\\
163  & $9207.458$ &  $35.624$\\
 $\vdots$&   $\vdots$ &  $\vdots$\\
245  & $13942.645$ &  $74.100$\\
246  & $14016.745$ &  $74.446$\\
247  & $14091.192$ &  $74.793$\\
$\vdots$&   $\vdots$ &  $\vdots$\\
\hline
\end{tabular}
\caption{The band structure of the reduced model is listed. 
The lower edge and band width are shown, respectively.
The sphaleron energy is about 9.08 TeV.}
\label{band_structure}
\end{table}
\paragraph{Thermal transition rate.---}
Now we scrutinize how much the band structure affects EWBG.
The typical energy scale of EWBG is a critical temperature $T_C$ 
at which the Higgs potential has degenerate minima if the electroweak phase transition is of first order,
which is about $\mathcal{O}(100)$ GeV in most cases. 

The decay rate of a false vacuum at finite temperatures is defined by~\cite{Affleck:1980ac} 
\begin{align}
\Gamma_A(T)=\frac{1}{Z_0(T)}\int^{\infty}_0dE~J(E)e^{-E/T}, \label{decay_rate}
\end{align}
where $Z_0(T)=[2\sinh (\omega_0/(2T))]^{-1}$ is a partition function of a harmonic oscillator with an angular frequency 
$\omega_0=g_2v\sqrt{V^{\prime\prime}(0)/M(0)}$ and $J(E)$ is a probability current, which is cast into the form $J(E)=T(E)/2\pi$.
The WKB approximation in high temperature regime of the field theoretic expression of (\ref{decay_rate}) is used to estimate 
the $(B+L)$-changing rate in electroweak theories~\cite{ArnoldMcLerran}.
We numerically calculate $\Gamma_A(T)$ for a single energy barrier in the region of $0\le \mu\le\pi$.


In the current framework of the periodic potential , we newly define the {\it transition rate} as
\begin{align}
\Gamma(T)=\frac{1}{Z_0(T)}\int^{\infty}_0dE~\frac{\eta(E)}{2\pi}e^{-E/T}, 
\label{Gam}
\end{align} 
where $\eta(E)$ is the density of states that embodies the band effect:
$\eta(E)=1$ for the conducting band and $\eta(E)=0$ for the band gap. 
It should be noted that  although what is really needed is the thermal average of the transition probability from $\mu=0$ to $\mu=\pi$,
the definition (\ref{Gam}) corresponds to setting the probability to unity for a state in one of the energy bands.
Therefore, the above naive definition of the rate provides an overestimated result. We defer the more precise estimate to future work.

\begin{figure}[t]
\center
\includegraphics[width=7cm]{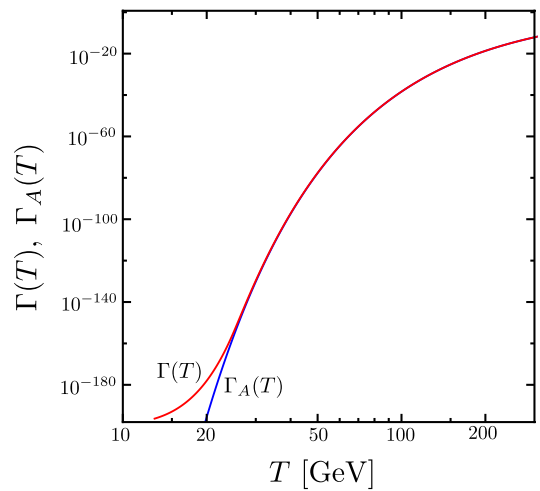}
\caption{$\Gamma(T)$ and $\Gamma_A(T)$ are shown against temperature,
where $\Gamma(T)$ is the transition rate taking the band effect into account while
$\Gamma_A(T)$ is the ordinary transition rate.}
\label{fig:R_T}
\end{figure}

The numerical values of $\Gamma(T)$ and $\Gamma_A(T)$ are plotted in Fig.~\ref{fig:R_T}.
Now we study the effect of the band structure on the BNPC, $\Gamma(T) < H(T)$, as needed for successful EWBG,
where $H(T)\propto  T^2/m_{\rm Pl}$ with $m_{\rm Pl}$ being the Planck mass is the Hubble parameter at $T$.
Although the enhancement due to the band structure is sizable at temperatures below 30 GeV, $\Gamma(T)$ is still too small to affect the BNPC quantitatively,
since $H(T)$ decreases more slowly than $\Gamma(T)$ for lower temperatures.
On the other hand, the enhancement appears very small near the electroweak phase transition temperature.
%
%
To quantify the effect, we denote $R(T) = \Gamma(T)/\Gamma_A(T)$, where $\Gamma_A(T)$ corresponds to the $(B+L$)-changing rate so far used.
The  modified BNPC is expressed as~\cite{Funakubo:2009eg}
\begin{align}
\frac{v(T)}{T}>\frac{g_2}{4\pi(\beta_1+\beta_2)}\bigg[42.97+\log{\cal N}+\log R(T)+\cdots \bigg],
\label{BNPC}
\end{align}
where ${\cal N}$ denotes the translational and rotational zero-mode factors of the fluctuations about the sphaleron, which may amount to about 10\% correction to the leading constant term 
(see, {\it e.g.}, Ref.~\cite{Funakubo:2009eg}).
With $R(T)$ obtained above, one finds that $\log R(T=100~{\rm GeV})\simeq 0.05$, so the band structure has little effect on the criteria (\ref{BNPC}) . 

\paragraph{(B+L)-changing process in high-energy collisions.---}
It is an interesting question whether $(B+L)$-changing process is visible at colliders
such as LHC and Future Circular Collider-$hh$ etc. 
Tye and Wong claim that it might be possible thanks to the unsuppressed transition probability 
between the topologically inequivalent vacua.
However, it is known that the creation of the classical configurations from the high-energy collision of
the two particles suffers from another type of exponential suppression 
other than the tunneling suppression discussed above. 
In Refs.~\cite{Funakubo:1991hm,Funakubo:1992nq}, 
the $(B+L)$-changing scattering amplitude is formulated by use of the complete set of the coherent state, which is dominated by
the bounce configuration describing a transition from one vacuum to another.
It is shown that, in the leading order of the WKB approximation, the overlap between the coherent state and the $n$-particle state produces a multiplicative  
factor $\bm{k}_1^2e^{-\pi |\boldsymbol{k}_1|/m_W}\bm{k}_2^2e^{-\pi |\boldsymbol{k}_2|/m_W}\cdots \bm{k}_n^2e^{-\pi |\boldsymbol{k}_n|/m_W}$ 
with $\boldsymbol{k}_i$ being a spatial momentum of the particle $i$.
Therefore, the overlap between the coherent state and the in-state composed of two particles whose total momentum is $E_{\rm sph}$ yields a 
suppression factor $\sim e^{-\pi E_{\rm sph}/m_W}\simeq 10^{-155}$, rendering $(B+L)$-changing process unobservably small.
We note in passing that existence of the suppression can also be shown using the complete set of the field eigenstates
instead of that of the coherent states.

Since each factor $\bm{k}^2e^{-\pi |\boldsymbol{k}|/m_W}$ has a sharp peak at $|\bm{k}|\sim m_W$, 
the above overlap suppression might be circumvented if the incoming two particles get scattered into multiple 
$W$ bosons where each has a momentum close to the $W$ boson mass, 
and then couple to the sphaleron all together. 
For this process to occur, about 80 $W$ bosons must be produced by the initial state scatterings, 
which would have a phase space suppression factor $\sim(1/(4\pi)^2)^{80}\simeq10^{-176}$, 
again preventing the $(B+L)$-changing process from being visible in high-energy collisions. 
Those types of the suppressions still remain regardless of the band structure, 
which is not properly discussed in Ref.~\cite{Tye:2015tva} (see also Ref.~\cite{Bachas:2016ffl}).
It should be emphasized that the above types of the suppressions do not exist at temperatures higher than the weak scale,
since so many particles whose momenta are of order $m_W$ are populated to have sizable overlap with the classical configuration.

\paragraph{Conclusions.---}
We have constructed the reduced model along the noncontractible loop connecting the classical vacua with different $B+L$ in a gauge-invariant manner,
and investigated the energy eigenstates of the mode in detail.
Based on the results, we have scrutinized to what extent the band structure can affect the EWBG scenario.
Our findings show that the BNPC is not virtually altered 
for $T\le T_C$. 

Before closing, a few comments are in order.
First of all, we make a remark on the formulation of the $(B+L)$-changing process.
The energy eigenstates obtained here are {\it not} those of $B+L$ or $\mu$.
If an in-state of a high-energy collision is an eigenstate of $\mu$ or a state localized at some $\mu$, it should be expressed as
a superposition of the energy eigenstates. 
Such a state should be peaked at some energy in the asymptotic region where $B+L$ is almost conserved.
The correct $(B+L)$-changing rate must be formulated by taking into account this situation.
Secondly, we have not taken thermal corrections into account in our analysis. In order to improve our results quantitatively,
the reduced model should be constructed starting from the field theoretic model with thermal corrections, as done in \cite{Funakubo:2009eg} 
to improve the precision of the BNPC. These issues will be dealt with elsewhere~\cite{FFS}.

\begin{acknowledgments}
K.~Funakubo is supported by JSPS KAKENHI Grant Number JP15K05057. 
K.~Fuyuto is supported by Research Fellowships of the Japan Society for the Promotion of Science for Young Scientists No.~15J01079.
E.S. is supported in part by the Ministry of Science and Technology of R. O. C. under Grant No. MOST 104-2811-M-008-011.
\end{acknowledgments}


\end{document}